# Phase Segregation Dynamics in Mixed-Halide Perovskites Revealed by Plunge-Freeze Cryogenic Electron Microscopy

Qingyuan Fan*[1,2,3], Yi Cui*[1], Yanbin Li*[1], Julian A. Vigil*[4,5], Zhiqiao Jiang[1,4], Partha Nandi[6], Robert Colby[7], Chensong Zhang[8], Yi Cui[1,2,9#], Hemamala I. Karunadasa[2,4#], Aaron M. Lindenberg[1,2,3#]

[1]Department of Materials Science and Engineering, Stanford University, Stanford, California 94305, United States
[2]Stanford Institute for Materials and Energy Sciences (SIMES), SLAC National Accelerator Laboratory, Menlo Park, California 94025, United States
[3]Stanford PULSE Institute, SLAC National Accelerator Laboratory, Menlo Park, California 94025, United States
[4]Department of Chemistry, Stanford University, Stanford, California 94305, United States
[5]Department of Chemical Engineering, Stanford University, Stanford, California 94305, United States
[6]Corporate Strategic Research, ExxonMobil Research and Engineering, 1545 US Highway 22 East, Annandale, New Jersey 08807, United States
[7]ExxonMobil Technology and Engineering Company, Annandale, New Jersey 08801, United States.
[8]Department of Molecular and Cellular Physiology, Stanford, Stanford, California, 94305, United States
[9]Department of Energy Science and Engineering, Stanford University, Stanford, California 94305, United States

*These authors contributed equally to this work
# Correspondence: A.M.L (email: aaronl@stanford.edu ), Y.C. (email: yicui@stanford.edu), H.I.K. (email: hemamala@stanford.edu )

## Abstract
Mixed-halide lead perovskites, with photoexcited charge-carrier properties suitable for high-efficiency photovoltaics, hold significant promise for high-efficiency tandem solar cells. However, phase segregation under illumination, where an iodide-rich phase forms carrier trap states, remains a barrier to applications. This study employs plunge-freeze cryogenic electron microscopy to visualize nanoscale phase segregation dynamics in $CsPb(Br_xI_{1-x})_3$ films. By rapidly freezing the illuminated samples, we preserve transient photoexcited ion distributions for high-resolution structural and compositional analysis at the nanoscale. Cryogenic scanning transmission electron microscopy techniques (EELS, 4D-STEM) captured the dynamics of photo-induced iodine migration from grain boundaries to centers, identified the buildup of anisotropic strain, and captured the heterogeneous evolution of this process within a single grain. These findings provide new insights into microscopic phase segregation mechanisms and their dynamics, enhancing our understanding of mixed-halide perovskite photostability.

## Main text
Halide perovskites have garnered interest in the field of optoelectronics due to the ease of large-area, low-cost film preparation methods and their high photoluminescence and photovoltaic conversion efficiency.[1,2,3,4] With a common chemical formula of $ABX_3$, various A-site cations (e.g., $Cs^+$, $CH_3NH_3^+$ ($MA^+$), $CH(NH_2)_2^+$ ($FA^+$)) and X-site anions (e.g., $I^-$, $Br^-$, $Cl^-$) have been widely investigated in the perovskites with $Pb^{2+}$ as the B-site cation. Among these materials, mixed-halide perovskites are particularly suitable for building tandem solar cells[5] since the bandgap is tunable from ultraviolet to near-infrared by varying the X-site (halide) composition.[6,7,8] However, photo-instability, manifesting as demixing of the halides under illumination, remains an important obstacle for their application in photovoltaic and light-emitting devices[9,10]. The photo-induced formation of iodide-rich domains induces undesired bandgap decreases, which results in decreased open-circuit voltages and reduced device-level performance.[8,11]

Nearly a decade has passed since Hoke, et al. first reported the observation of phase segregation in $(CH_3NH_3)Pb(Br_xI_{1-x})_3$,[12] but the fundamental microscopic mechanisms underlying this behavior, and even basic aspects of how the local stoichiometry changes, remain controversial and widely debated[10]. Bulk characterization methods, such as photoluminescence (PL) and X-ray diffraction (XRD),[12] were first employed to verify and study the extent of phase segregation. Upon visible-light illumination, the red-shifted PL emission and split X-ray diffraction peaks indicated the formation of phases with distinct halide ratios.[12,13] Various mechanisms have been proposed, typically involving non-equilibrium dynamical effects, to explain the initiation and evolution of ion migration. For example, the proposed driving forces for phase segregation include the temperature-dependent miscibility gap,[14] the release of polaronic strain through the formation of I-rich domains,[15] carrier funneling towards lower-bandgap I-rich regions,[16,17] and defects, especially near grain boundaries.[18]

Spatially resolved optical probes, such as confocal photoluminescence[13,18,19,20] and cathodoluminescence,[15] have been employed to identify regions with different halide compositions.[12] Kelvin probe force microscopy has revealed the contact surface potential, with sensitivity to ion migration at grain boundaries.[21] To achieve more direct elemental characterization, techniques such as energy-dispersive X-ray spectroscopy (EDX, EDS)[17] and electron energy loss spectroscopy (EELS) offer detailed information on the spatial distribution of I-rich and Br-rich regions. The majority of studies state that the distributions of I-rich and Br-rich phases are influenced by morphology, with many specifically indicating that I-rich phases tend to form near grain boundary and sample edges.[12,19,18] However, there are also reports of I-rich and Br-rich phases forming across grain boundaries or of I-rich domains appearing in the grain centers, away from grain boundaries, or in single crystals.[22,23] These contradictory findings highlight the need for in-situ microscopic studies and high-resolution analysis of defects and strain states as the phase segregation process occurs.

A central challenge in elucidating the mechanisms for photodegradation arises from the intrinsically dynamical nature of these processes. Phase segregation is reversible over minutes to hours without illumination, and measurements which average in space or time essentially blur the atomic to mesoscale processes that are occurring. Prior spatially resolved methods have not been able to provide an in-situ view of this process or to track the underlying intrinsically dynamical ion motions under device-relevant conditions. The scanning-based PL and CL measurements usually take minutes or longer to complete, especially when the spatial resolution is high. As the result, these techniques are usually applied after illumination or at equilibrium under continuous illumination, without the ability to capture the transient moments of the segregation process[17]. Further, optical probe methods often provide only indirect measurements, relying on local bandgap variations in the material. Due to the carrier funneling effect[13,24] and the resulting significantly higher PL efficiency in I-rich domains, PL imaging typically highlights regions with the highest iodide composition, potentially overlooking areas where iodide is not predominant. Similarly, cathodoluminescence measurements are operated under electron irradiation in vacuum, and provide indirect measurements based on the optical properties of the material. Therefore, characterization approaches that offer high spatial resolution while simultaneously capturing the transient ion distribution and its evolution are crucial to understanding the photodegradation mechanisms.

Here we apply for the first time an approach that enables the tracking of the transient ion distribution and nanoscale structural changes induced by illumination at room temperature. This method utilizes a plunge-freeze technique[25] (Fig. 1), where photoexcitation at room temperature under normal atmospheric conditions triggers the segregation process, which then develops over a programmable time period. The samples are then rapidly plunged into liquid nitrogen, effectively freezing the transient configuration, which is then analyzed using cryo-electron microscopy techniques. The cryogenic conditions also address the issue of mixed-halide perovskites' sensitivity to electron doses at room temperature[25,26].

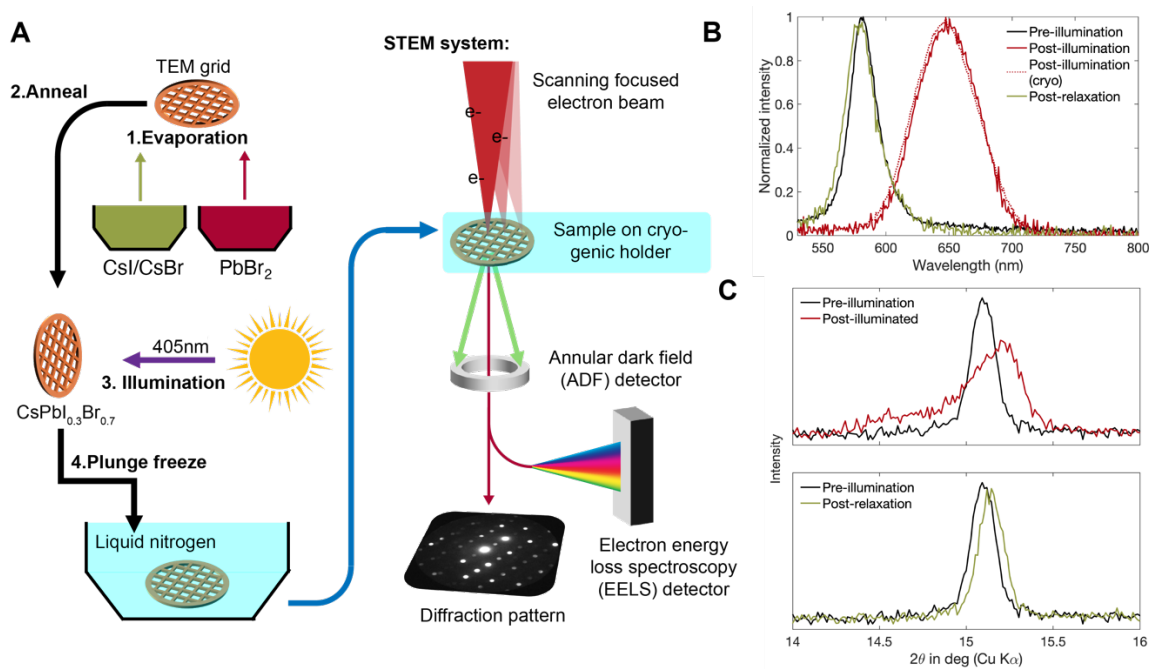

**Figure 1. Characterization approach for capturing phase segregation.** (A) Schematic diagram of the sequential sample preparation, plunge-freeze approach, and coupled cryo-electron microscopy approach reported here. Photoluminescence emission spectra (excitation wavelength = 405nm) (B) and thin-film XRD patterns measured in the region of the (100) pseudo-cubic Bragg reflection (C), including for pristine $CsPb(Br_{0.7}I_{0.3})_3$ (pre-illumination), post-illumination, post-illumination (cryo) stored for 3 weeks, and post-relaxation (see main text for description of each condition).

Here, we prepared $CsPb(Br_xI_{1-x})_3$ perovskite films with thicknesses of ca. 100 nm by sequential thermal evaporation of $CsI/CsBr$ and $PbBr_2$ directly on carbon-film-Cu TEM grids, followed by annealing under $N_2$ atmosphere (see Supporting Information for Experimental Methods). PL emission and XRD measurements, averaged over the entire sample, are conducted to verify the bandgap and pseudo-cubic perovskite structure, respectively, of the pristine polycrystalline film. To trigger photo-induced phase segregation at room temperature, the sample was illuminated under $N_2$ atmosphere using a 405-nm LED, with an intensity of 0.4 W/cm². The illumination is conducted within the plunge-freeze setup (Fig. 1A) to freeze the sample once a desired time of illumination is reached, allowing for subsequent characterization of the frozen-in structure. We note that symmetry-lowering solid-solid phase transitions are well known in the perovskites upon cooling from high temperature [27], however rapid freezing helps prevent phase transitions that might occur under slower, equilibrium cooling. Particular to this work, the equilibrium phase diagram for

CsPb(Br$_x$I$_{1-x}$)$_3$ has primarily been investigated at high temperatures (e.g. 330 °C, where the phase is cubic for all $x$)[28]; at room temperature, most compositions within CsPb(Br$_x$I$_{1-x}$)$_3$ appear stable as a polycrystalline film[29], though the descent in symmetry from the high-temperature cubic phase differs with $x$, growth and annealing conditions, and the substrate. Thus, we report all Bragg reflections with (pseudo)cubic Miller indices for consistency without a discernable tetragonal or orthorhombic distortion, which indicates the absence of phase changes during the plunge-freeze operations.

The PL emission spectrum was first measured following various illumination times, ranging from 0 to 8 minutes. Prior to illumination, the PL peak at 580 nm (Fig. 1B, pre-illumination) indicates a halide composition ratio of I/Br = 0.45, derived by comparing the PL spectrum with those of previously studied perovskites with varying halide compositions[32]. EELS characterization, discussed later, illustrates that the composition is CsPb(Br$_{0.7}$I$_{0.3}$)$_3$. Upon illumination, the PL peak redshifts and the intensity increase begins to saturate at 5 minutes, indicating that the phase segregation process is fully achieved after 5 minutes of illumination (Fig. S7, Supplemental Information). After illumination for 5 minutes, the PL peak redshifts to 651 nm (Fig. 1B, post-illumination), matching the PL wavelength of CsPb(Br$_{0.27}$I$_{0.73}$)$_3$,[30] and indicating the formation of a lower-bandgap I-rich phase. The same measurement is applied to the sample illuminated for 5 minutes, followed by 2 hours of resting in a dark environment. The post-relaxation PL spectrum (Fig. 1B) shows a peak wavelength of 579 nm, which demonstrates the reversibility of the photo-induced phase segregation. A small blue shift is likely due to partial iodide loss under illumination, as the lower iodide ratio corresponds to a smaller PL emission wavelength.

On the same samples, XRD results provide structural information that corroborates the PL data. The thin-film XRD pattern shows strong ($h$00) pseudo-cubic Bragg reflections from the pristine CsPb(Br$_{0.7}$I$_{0.3}$)$_3$ structure (Fig. S5, Supplemental Information), with a corresponding pseudo-cubic lattice parameter of ca. 5.87 Å. In Fig. 1C, top, the (100) Bragg peak exhibits broadening after illumination, indicating a wider lattice-parameter distribution concomitant with phase segregation. The similarity of the XRD patterns following relaxation (compared to the pristine film) also demonstrates reversibility in the average structure, albeit with a small, irreversible lattice-parameter decrease, again likely arising from iodide loss.

To carry out high-resolution cryogenic characterization, photoexcited samples are subsequently plunge-frozen at $T = 77$ K. For post-illumination samples, the PL spectrum measured after three weeks of storage in liquid nitrogen (Fig. 1B, red dashed curve) is in good agreement with that obtained at room temperature (Fig. 1B, red solid curve), indicating that cryogenic conditions inhibit reversible ion migration while preserving the crystal structure and frozen in, transient ion distribution. The plunge-freeze approach thus provides access to the intrinsic, instantaneous configuration of the photoexcited, phase-segregated samples.

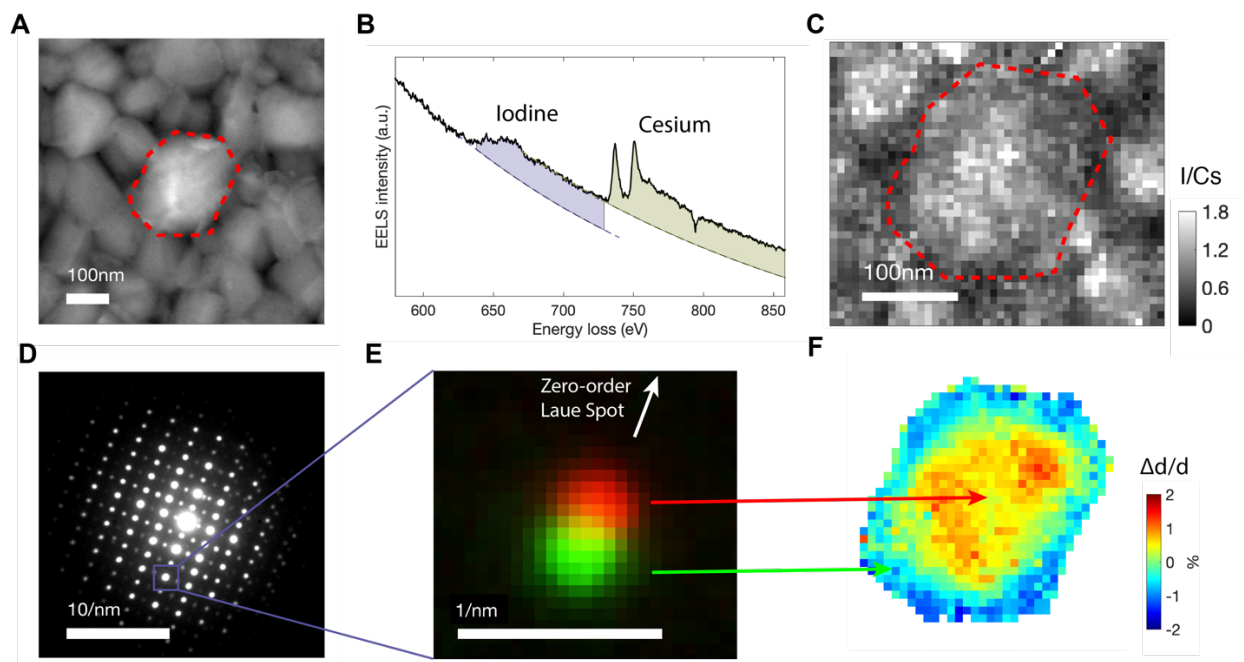

**Figure 2. Morphology, stoichiometry, and structural imaging of the photoexcited mixed-halide perovskite.** (A) Annular dark field (ADF) image of the polycrystalline sample. The grain in the center, outlined by the red dashed profile, indicates the area analyzed in subsequent EELS (C) and 4D-STEM (F) results. (B) Representative EELS spectrum of an area on the grain. The color filling indicates the integration area for calculating the areal atomic density of iodine (blue) and cesium (green). (C) Mapping of the I/Cs ratio (integrated counts) over the area for frozen configuration at $t = 5$ minutes illumination. Each pixel includes calculation as shown in (B). (D) Selected area diffraction pattern of a 40nm x 40nm area in the grain center. (E) The diffraction peaks (red: in the grain center, green: near grain boundary) zoomed in to the purple square in (D). (F) Map of percent lattice-parameter change ($\Delta d/d$) at $t =$ 5 minutes illumination. The value for each pixel is calculated from the diffraction pattern, averaging over several diffraction peaks. The pixel size is 10 nm × 10 nm.

To explore the phase segregation process further, three cryogenic STEM-based techniques—annular dark-field imaging (ADF), electron energy loss spectroscopy (EELS), and 4D-STEM—are employed to analyze the same sample areas, providing insights into morphology, stoichiometry, and structure, respectively. The ADF images reveal variations in grain size ranging from 100 nm to 400 nm (Fig. 2A).

In this study, EELS provides areal atomic density data for the elements. The EELS spectrum (Fig. 2B), particularly the $M_5$ peak for I and the $M_4$ peak for Cs, is used to calculate the areal atomic density of these elements for each pixel across the imaging area. Since the EELS detector's range does not encompass the energy of Br peaks, a direct calculation of the ratio between I and Br is not feasible. The atomic areal densities of iodine are influenced by the sample thickness, which is smaller at the grain boundary. Since the focus is on the stoichiometric composition of halide elements rather than areal atomic density, it is necessary to remove the influence of sample thickness to accurately represent the halide ratio distribution. Diffraction patterns (Fig. 2D and Fig. S7, Supplemental Information) confirm that the perovskite structure remains unchanged, suggesting that the volume atomic density of $Cs^+$ is constant. Therefore, the areal density of cesium is proportional to the sample thickness. By calculating the areal atomic density ratio of iodine to cesium (I/Cs) within the perovskite, the halide composition can be accurately determined. This

I/Cs ratio is mapped with a resolution of 10 nm and displayed in Fig. 2C, corresponding to the transient elemental configuration after photoexcitation for t = 5 minutes (discussed further below).

4D-STEM provides complementary high-resolution local structural information.[31] Utilizing the STEM mode of a cryogenic microscope, a focused electron beam with a small divergence angle generates a 2D diffraction pattern at each detected sample position. By scanning the electron beam across the sample surface, detailed structural information is obtained. This technique not only reveals crystal symmetry and orientation, but also enables quantitative analysis of the average lattice parameter via reciprocal-space mapping. In this study, diffraction patterns and quantitative analyses are performed at each pixel within the scanned area, with a pixel resolution of 10 nm.

The single-crystal diffraction patterns obtained from the 4D-STEM results (an example shown in Fig. 2D) confirm that the crystal remains in the (pseudo)cubic perovskite phase upon cooling, with a lattice parameter of 5.80 Å at $T$ = 77K. Furthermore, diffraction patterns taken at various positions within a grain confirm that I-rich and Br-rich domains within the same grains maintain a continuous crystal structure, sharing the same space-group symmetry and local orientation.

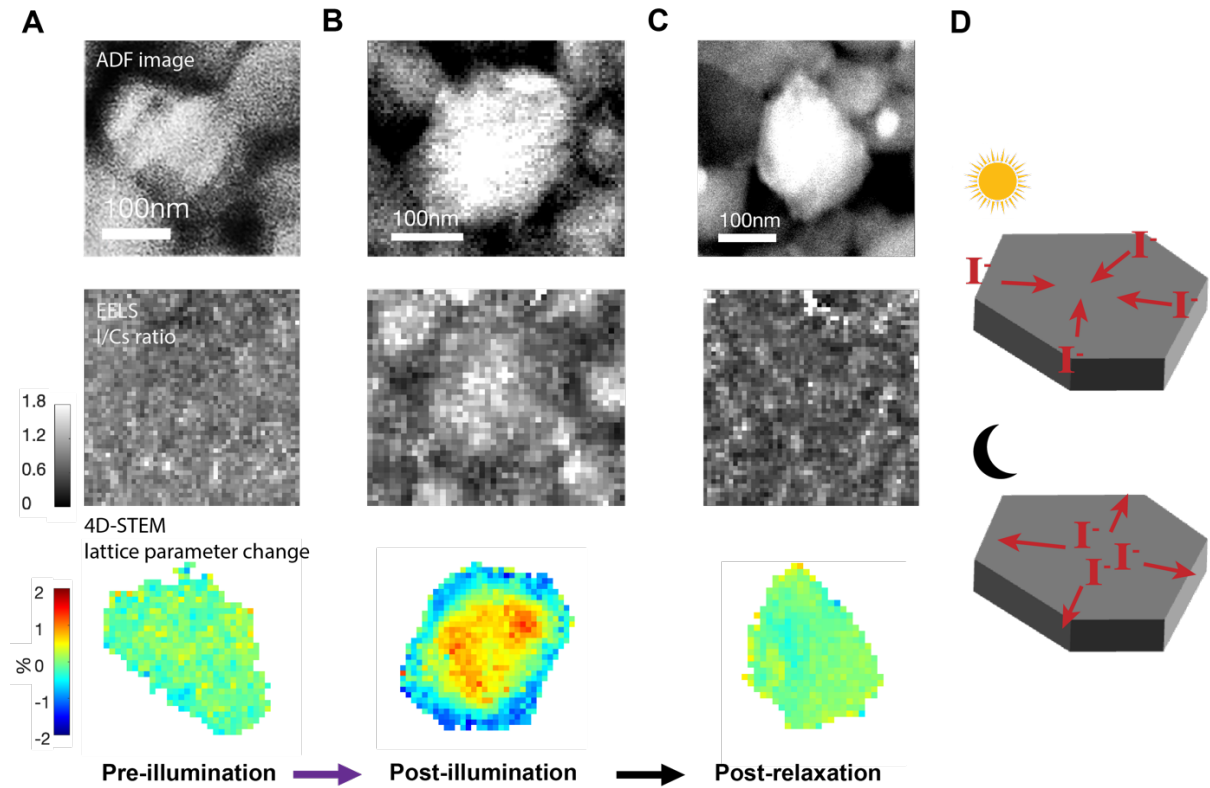

**Figure 3. Correlated morphology, stoichiometry and structural imaging for the pristine, illuminated, and relaxed, self-healed mixed-halide perovskite.** ADF image (top), I/Cs ratio from EELS mapping (middle), and percent lattice-parameter change from 4D-STEM (bottom), for (A) pre-illumination, (B) post-illumination (t = 5 minutes) and (C) post-relaxation samples. (D) Schematic diagram showing the apparent intragrain iodide migration directions while under illumination (top), and during dark relaxation (bottom).

The dynamical segregation process is studied by applying the EELS and 4D-STEM methods on the pre-illumination, post-illumination (t = 5 minutes) (corresponding to frozen in structure), and post-relaxation samples. The I/Cs ratio mapping shows uniform halide mixing in the pristine $CsPb(Br_{0.7}I_{0.3})_3$ perovskite samples (Fig. 3A). Post-illumination analysis shows distinct contrasts in the I/Cs ratio, with a higher iodine concentration at the grain center and lower concentrations near the grain boundaries, as revealed by correlating EELS mapping with ADF imaging (Fig. 3B). The EELS results from the post-relaxation samples (Fig. 3C) also demonstrate uniform iodine distribution, confirming the reversible nature of the phase segregation process.

After illumination in the frozen-in configuration, the averaged I/Cs ratio in the central part of the grains (I/Cs = 1.19) exceeds the averaged value in the pristine samples (I/Cs = 0.88), indicating iodine ion migration from near the grain boundaries towards the centers (schematically shown in Fig. 3D). The distribution of measured I/Cs ratios in the region within 30 nm of the grain boundary (blue) and that in the grain center (red) are shown by the histograms in Fig. 4. In addition to the differences in the average halide composition between these regions, Fig. 4 (middle) highlights the heterogeneity of the I/Cs ratio, revealing a wide distribution at both the grain boundary and grain center. We note that an average I/Cs ratio of 1.19 is expected to produce significantly smaller PL shifts, corresponding to emission at approximately 605 nm,[32] which contrasts with the shift to 651 nm observed experimentally (Fig. 1B). This discrepancy can be attributed to the long tails in the distribution of the I/Cs ratio, extending up to 1.8 (Fig. 4, middle top). These tails indicate that regions with high I/Cs ratios, though only a small fraction of the sample, exert a disproportionate influence on the PL shifts. This occurs because excited carriers tend to funnel into these regions with the lowest bandgap, thereby amplifying the observed PL shifts.[12,24, 9] 4D-STEM measurements of the plunge-frozen samples are correlated with ADF and EELS mapping. Although regions within the same grain exhibit identical crystal symmetry and orientation, variations in lattice parameter are observed. In Fig. 2E, diffraction peaks from different positions within a grain (red: center, green: near boundary) are plotted together for comparison under the same view box. The variation in peak position along the direction pointing to the zero-order diffraction spot indicates strain across the segregated grain area. The lattice parameter mapping (Fig. 3) derived from 4D-STEM, averaged over multiple diffraction peaks, shows trends that align with the EELS results. The uniform lattice parameter distribution across the pre-illumination grain matches the uniform I/Cs ratio. In the post-illumination samples, the regions with a large lattice parameter are predominantly located at the grain centers, a pattern that is reversed when samples are rested in a dark environment. Considering that iodine atoms are larger than bromine atoms, perovskites with higher iodine content exhibit larger lattice parameters, thus the results of the structural studies closely match the measurements on film stoichiometry. Additionally, the 4D-STEM results reveal no pre-existing strain within the detection limits before phase segregation, whereas the post-illumination strain distribution shows values of over 4%.

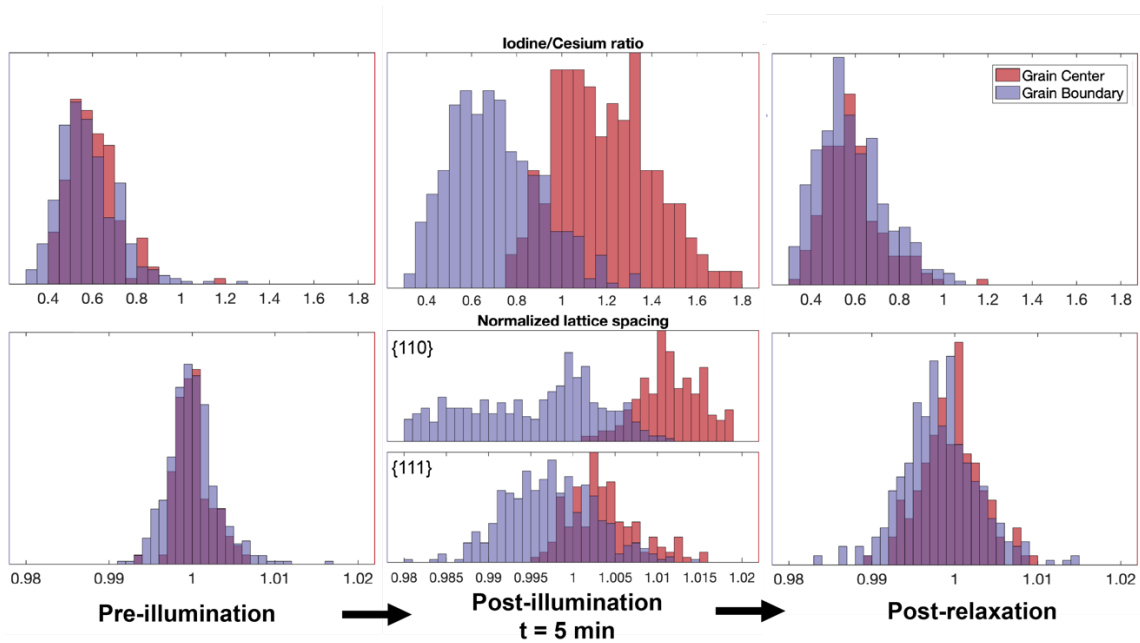

**Figure 4. Histograms of the measured I/Cs ratio (EELS) and strain (4D-STEM) of various samples and areas.** Histogram of the I/Cs ratio (top) and normalized lattice-parameter change (bottom) for pre-illumination, post-illumination and post-relaxation samples. The blue histograms (grain boundary) count the data from any area less than 30 nm away from the grain boundary, indicating the near-boundary area. The red (grain center) histograms count the data from any area more than 30 nm away from the grain boundary. The lattice parameters for pre-illumination and post-relaxation samples are analyzed by averaging the lattice parameters corresponding to all diffraction peaks, while the lattice parameters are separated by two orthogonal families of planes ({110} and {111}) on the post-illumination samples to illustrate the anisotropic strain.

Fig. 4 (bottom) illustrates the histograms of lattice parameter distribution, which are calculated by averaging the lattice parameters corresponding to multiple diffraction peaks. For the pre-illumination and post-relaxation samples without phase segregation, the lattice parameter distributions are narrow. The lattice parameters of the grain center (red) and grain boundary (blue) areas show a similar lattice with a narrow distribution, which corresponds with the uniform I/Cs stoichiometry. In contrast, the lattice parameter distribution of the post-illumination (5 minutes) films are broader, with different averaged value for grain center and grain boundary samples, which corresponds with the heterogeneous I/Cs ratio for the samples with segregation. Here the lattice parameter distribution is analyzed separately for two orthogonal families of planes (top:{110}, bottom:{111}). The histograms in Fig. 4 (bottom, middle) show a larger lattice parameter contrast between grain center and near-boundary areas for {110} planes, while the contrast is less pronounced in the {111} planes indicating anisotropic strain. The two families of planes are labeled in Fig. 5B, with the spatial mapping of lattice parameters shown in Fig. 5C-D.

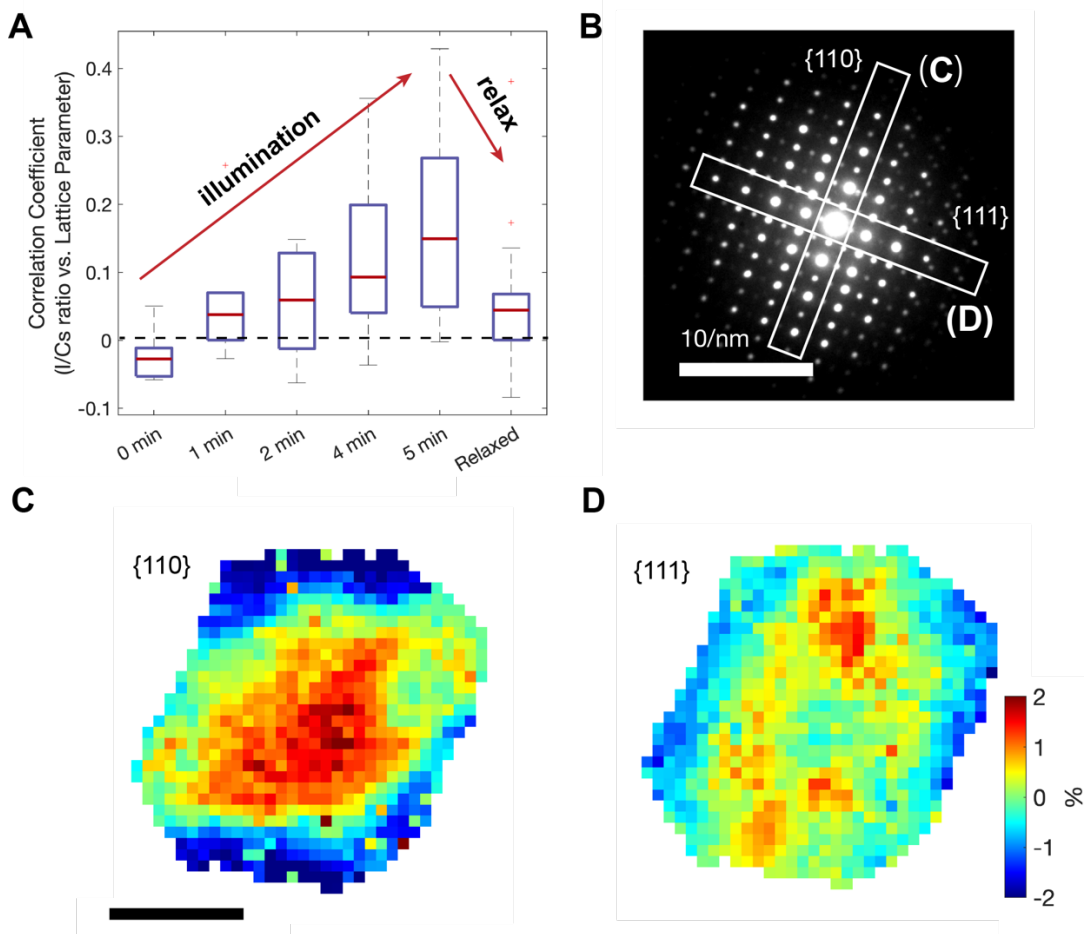

**Figure 5. Evolution and anisotropy of the intragrain photo-induced phase segregation process**. (A) Pearson product-moment correlation coefficients between ADF and EELS mapping (I/Cs ratio) with various illumination time and post-dark relaxation at the grain center. (B) Selected area diffraction pattern. The percentage change in lattice parameter—calculated from diffraction families shown in (C) and (D)— are mapped using the same color palette.

To quantitatively characterize the morphology-dependence and time-dependence of the I-rich domain distribution, the correlation coefficient between the ADF signal and the I/Cs ratio is calculated for various light exposure times (Figure 5A). The ADF signal is higher at the grain center and lower at the grain boundary, a distribution that correlates with the thickness difference in these regions. Thus, a positive correlation coefficient corresponds to a higher iodine composition at the grain center. The pristine sample, which has a uniform I/Cs ratio, yields coefficients close to zero, whereas the segregated samples exhibit positive correlation values that grow in with cumulative illumination time. In particular, data from over 80 EELS and ADF mappings on samples with varying illumination times reveal an increasing iodide concentration at the grain center. The correlation coefficients between the I/Cs ratio and ADF values show a gradual increase at the onset of segregation, with illumination times as short as 1 and 2 minutes. These correlation results suggest that the ion migration process is morphology-dependent from the early stages of segregation, challenging the notion that I-rich clusters nucleate stochastically before migrating to specific regions of the grain (center/boundary). [22] The broadening of the correlation distribution with time, as shown by the error bars in Fig. 5A, also indicates the key role of heterogeneity in this process, consistent with the histograms shown in Fig. 4.

The 4D-STEM measurements additionally allow for analysis of the anisotropic nature of the phase segregation process within a single grain. Distinct from the maps shown in Fig. 3C, the lattice parameter derived from two orthogonal families of planes ({110} and {111}) are mapped in Fig. 5C and D, respectively, exhibiting significantly larger contrast shown in Fig. 5C as compared to Fig. 5D. This observation indicates segregation-induced anisotropy within grains, with strain being larger along certain local crystallographic directions. A more comprehensive analysis of anisotropy, including a correlation coefficient mapping that tracks individual diffraction peaks, is presented in the Supplementary Information (Fig. S4). The spatial distribution of the lattice parameter also suggests a directional lattice strain, extending from the grain center to the boundary. This anisotropic strain may be determined by the grain morphology or other pre-existing strain or defect conditions. Understanding the anisotropy of the strain within individual grains is important for understanding of the phase segregation mechanism, including the influence of initial morphology and defects.

Previous reports have documented various spatial distributions of I-rich regions during phase segregation with contradictory results[33,34]. The majority of the literature supports the conclusion that phase segregation tends to occur near grain boundaries[15,20,21,17] or crystal edges[13,18], though some also demonstrated boundary-independence in large grain samples[22]. Specifically, many studies have argued that I-rich domains often form near these boundaries[13,18], contrary to our findings. Others have observed that I-rich and Br-rich domains form across the boundaries[17]. Additionally, certain PL mapping studies have identified I-rich clusters or phases at the grain centers in all or some of the samples[23,15], which contradicts previous results but is consistent with our measurements. So far, it is not clear whether these findings represent an intrinsic and universal property of the phase-segregation mechanism or if the microscopic details are strongly influenced by defect chemistry, substrates, grain size, strain, or fabrication method. High spatial resolution characterization approaches as presented here hold promise to explore dependencies on spatial defects. Overall, this research indicates via direct measurements that dynamic iodine motion occurs towards the grain center under photoexcitation and identifies the heterogeneity in the phase segregation process.

**Conclusion**. This research provides a new window into phase segregation in mixed-halide perovskite films, specifically $CsPb(Br_{0.7}I_{0.3})_3$, using cryogenic electron microscopy combined with a plunge freeze technique, focusing on the spatial distribution of I-rich and Br-rich domains within the perovskite structure. The cryogenic approach effectively preserves the transient ion distributions, offering a new view of the dynamics and reversible nature of light-induced phase segregation while maintaining the structural integrity of the material. The use of advanced electron microscopy techniques, including ADF, EELS, and 4D-STEM, provides high-resolution insights into the morphological, compositional, and structural details of the segregated phases. This research shows the time-dependent development of a higher concentration of iodine at the grain centers from element density mapping, supported with strain mapping. The usage of 4D-STEM also shows the anisotropic and heterogeneous strain after segregation for the first time. Finally, the development of the plunge-freeze method also defines new opportunities in the characterization of the dynamical properties of the halide perovskites and other materials undergoing light-induced structural transitions and other thermally activated processes.

## EXPERIMENTAL PROCEDURES:

### Sample preparation

See Supporting Information.

### Illumination and plunge-freeze

The sample is mounted on a piston connected with compressed gas, as indicated in the supplementary information. A 405-nm UV LED is used for illumination with an intensity of 0.4 W/cm$^2$. The illumination time ranges from a half a minute to more than 8 minutes. After the desired illumination time, a value is opened for high-pressure compressed gas to quickly drop the sample to the liquid nitrogen container to freeze the sample. The operations afterwards are at cryogenic temperature, either in liquid nitrogen or in vacuum to avoid the influence of moisture or ice.

### Cryogenic-STEM based characterization

The ADF imaging, EELS mapping and 4D-STEM are performed with a FEI 80-300 kV environmental transmission electron microscope (ETEM) at 300 kV. The CsPb(Br$_x$I$_{1-x}$)$_3$ sample on TEM grids operated by illumination is transferred to Gatan 636 Single Tilt Liquid Nitrogen Cryo Transfer Holder under liquid nitrogen. The sample is kept at $T = 98$ K through the whole process of characterization. Firstly, the morphology image is obtained with the ADF detector. Then EELS mapping and 4D-STEM are measured correlatedly at the same area/grain. Then the 4D-STEM measurement is taken at low dose rate ~4e/A$^2$s as measured from the fluorescent screen, with exposure time of 0.04s for each pixel. The diffraction pattern for each pixel is extracted to show the crystal orientation. In these diffraction patterns, the positions of each diffraction peak, as well as the zero-order peak, is fitted with Gaussian function. The lattice-parameter mapping is achieved by averaging the corresponding {100} plane spacing and calculating ($d_{hkl} = \frac{a}{\sqrt{h^2+k^2+l^2}}$) the lattice parameter at each pixel. Then the same area is scanned again with the EELS detector, with electron dose of <20 e/A$^2$s and exposure time of 0.5s to avoid damage. The EELS detector measures the electron counts at the range at 569eV – 1081 eV, covering the range of M5 peak for iodine at 619 nm and M4 peak of cesium at 726 nm. The integral over the EELS peak calculate the areal atomic density of I and Cs at each pixel of scanning, which clearly shows the thickness variation separately and is used for calculating the I/Cs ratio.

## ACKNOWLEDGEMENTS:

This work was supported by Exxon Mobil through its membership in the Stanford Strategic Energy Alliance. Y.C. acknowledges the cryo-EM technique development support from the US Department of Energy, Office of Basic Energy Sciences, Division of Materials Science and Engineering under contract DE-AC02-76SF00515. J.A.V. acknowledges fellowship support from the Stanford University Office of the Vice Provost of Graduate Education and the NSF Graduate Research Fellowship Program under Grant No. DGE-1656518.

**AUTHOR CONTRIBUTIONS:**

A. M. L., H. I. K. and Prof. Y.C. conceived the idea. J.A.V. and Z.J. synthesized the perovskite samples and collected XRD measurements. Q.F., Y.C. and Y.L. performed the optical operations and plunge-freezing of samples, as well as the cryogenic microscopic characterizations. Y.L. conducted the TEM imaging analysis. C. Z. helped on the cryogenic operations of the samples. Q.F. conducted the analysis of the EELS and 4D-STEM results. Q.F., Y.C. and Z.J. co-wrote the original manuscript. All authors discussed the results and commented on the manuscript.

**DECLARATION OF INTERESTS:**

The authors declare no conflict of interest.